\documentclass[prc,superscriptaddress,unsortedaddress,twocolumn,showpacs,preprintnumbers,amsmath,amssymb,floatfix,dvipdfmx]{revtex4}
\usepackage{amsmath}
\usepackage{amssymb}
\usepackage{times}
\usepackage{mathrsfs}
\usepackage{multirow}
\usepackage{bm}
\usepackage{ulem}
\usepackage{color}
\usepackage[dvipdfmx]{graphicx}

\def\la{\mathrel{\mathpalette\fun <}}
\def\ga{\mathrel{\mathpalette\fun >}}
\def\fun#1#2{\lower3.6pt\vbox{\baselineskip0pt\lineskip.9pt
\ialign{$\mathsurround=0pt#1\hfil##\hfil$\crcr#2\crcr\sim\crcr}}}

\newcommand{\beq}{\begin{equation}}
\newcommand{\eeq}{\end{equation}}
\newcommand{\bea}{\begin{eqnarray}}
\newcommand{\eea}{\end{eqnarray}}

\newcommand{\bfi}[1]{\mbox{\boldmath $#1$}}
\newcommand{\bfis}[1]{\mbox{\boldmath ${\scriptstyle #1}$}}

\newcommand{\vk}{{\bfi k}}

\newcommand{\vrr}{{\bfi r}}
\newcommand{\vR}{{\bfi R}}

\newcommand{\vik}{{\bfis k}}

\def\bk{\mbox{\boldmath $k$}}

\begin{document}

\title{
Microscopic calculations based on chiral two- and three-nucleon forces \\
for proton- and $^{4}$He-nucleus scattering}

\author{Masakazu Toyokawa}
\email[]{toyokawa@phys.kyushu-u.ac.jp}
\affiliation{Department of Physics, Kyushu University, Fukuoka 812-8581, Japan}

\author{Masanobu Yahiro}
\affiliation{Department of Physics, Kyushu University, Fukuoka 812-8581, Japan}

\author{Takuma Matsumoto}
\affiliation{Department of Physics, Kyushu University, Fukuoka 812-8581, Japan}

\author{Kosho Minomo}
\affiliation{Research Center for Nuclear Physics (RCNP), Osaka
University, Ibaraki 567-0047, Japan} 

\author{Kazuyuki Ogata}
\affiliation{Research Center for Nuclear Physics (RCNP), Osaka
University, Ibaraki 567-0047, Japan}

\author{Michio Kohno}
\affiliation{Research Center for Nuclear Physics (RCNP), Osaka
University, Ibaraki 567-0047, Japan}

\date{\today}

\begin{abstract} 
We investigate the effects of 
chiral three-nucleon force (3NF) on proton scattering at 65 MeV 
and $^{4}$He scattering at 72 MeV/nucleon from heavier targets, 
using the standard microscopic framework composed of 
the Brueckner-Hartree-Fock (BHF) method and the $g$-matrix folding model. 
For nuclear matter, the $g$ matrix is evaluated  
from chiral two-nucleon force (2NF) of N$^{3}$LO and chiral 3NF of NNLO  
by using the BHF method. Since the $g$ matrix thus obtained is numerical 
and nonlocal, an optimum local form is determined 
from the on-shell and near-on-shell components of $g$ matrix 
that are important for elastic scattering. 
For elastic scattering, 
the optical potentials  are calculated by folding 
the local chiral $g$ matrix with projectile and target densities. 
This microscopic framework reproduces the experimental data 
without introducing any adjustable parameter. 
Chiral-3NF effects are small for proton scattering, 
but sizable for $^{4}$He scattering at middle angles 
where the data are available. 
Chiral 3NF, mainly in the 2$\pi$-exchange diagram, 
makes the folding potential less attractive and more absorptive 
for all the scattering.  
\end{abstract}


\maketitle

\section{Introduction}
\label{Introduction}
An important current issue in nuclear physics 
is to understand the effects of 
three-nucleon force (3NF) on finite nuclei, nuclear reactions and 
nuclear matter. 
Quantitatively decisive roles of 3NFs have been established in 
properties of light nuclei as well as of nuclear matter 
\cite{Hammer13}.
This issue started with the 2$\pi$-exchange 3NF 
proposed by Fujita and Miyazawa~\cite{Fuj57}. 
Recently, a major breakthrough was made on this issue with 
chiral effective field theory (Ch-EFT); 
see Refs.~\cite{Epelbaum-review-2009,Machleidt-2011} and references therein. 
Ch-EFT is a theory based on chiral perturbation theory to provide 
a low-momentum expansion of two-nucleon force (2NF), 3NF 
and many-nucleon forces. 
Using this theory, one can define multi-nucleon forces systematically. 
The effects of chiral 3NF were analyzed in many papers; e.g., see 
Ref.~\cite{Kalantar-2012} 
for light nuclei, 
Refs.~\cite{Holt14,Ekstrom:2015rta} 
for {\it ab initio} nuclear-structure calculations in lighter nuclei 
and Refs.~\cite{HEB11,Koh13,Kru13,Dri14,Kru15}
for nuclear matter. 
Recently the role of chiral four-nucleon forces was also analyzed for 
nuclear matter ~\cite{Kai12}.
When the $g$ matrix (the effective nucleon-nucleon interaction 
in nuclear medium) is calculated from chiral 2NF+3NF with 
the Brueckner-Hartree-Fock (BHF) method, it   
well accounts for the empirical properties 
of symmetric nuclear matter~\cite{Koh13}. 
The $g$ matrix depends on the nuclear-matter density $\rho$; i.e.,  
$g=g(\rho)$. Chiral-3NF effects become more important 
as $\rho$ increases.

Another important issue in nuclear physics is microscopic understanding 
of nucleon-nucleus (NA) and nucleus-nucleus (AA) optical potentials. 
The optical potentials are essential in describing not only 
elastic scattering but also inelastic scattering 
and transfer and breakup reactions. In fact, 
the optical potentials are essential inputs in 
distorted-wave Born approximation and 
continuum discretized coupled-channel method (CDCC) calculations 
\cite{CDCC-review3}.

The $g$-matrix folding model is a standard method of calculating 
the optical potential microscopically. Actually, many 
nuclear reactions have been analyzed with the model. In the model, 
the potential is obtained by folding $g(\rho)$ with projectile and target 
densities ($\rho_{\rm P}$ and $\rho_{\rm T}$) for AA scattering
and with $\rho_{\rm T}$ for NA scattering; e.g., 
see Refs.~\cite{Brieva-Rook,Amos,CEG07,Raf13,MP} 
for $g(\rho)$ and 
Refs.~\cite{DFM-standard-form,Sum12,Egashira:2014zda} 
for the folding procedure. 
The model is called the single-folding (SF) model for NA scattering 
and the double-folding (DF) model for AA scattering.

For NA elastic scattering, the SF model based on 
the Melbourne $g$ matrix~\cite{Amos}, constructed from the Bonn-B 
nucleon-nucleon (NN) interaction \cite{BonnB}, 
well reproduces the experimental data with no adjustable parameter. 
In the folding procedure, 
the value of $\rho$ in $g(\rho)$ is assumed to be a value of $\rho_{\rm T}$ 
at the midpoint $\vrr_{\rm m}$ of interacting two nucleons: 
$\rho=\rho_{\rm T}(\vrr_{\rm m})$. 
Target-excitation effects on the elastic scattering 
are thus well described by the SF model based on 
the local-density approximation.

We have recently investigated chiral-3NF effects on the description 
of NA scattering \cite{Toyokawa:2014yma} and AA scattering 
\cite{Minomo:2014eqa}, $^{12}$C$+^{12}$C and $^{16}$O$+^{16}$O, 
by modifying the Melbourne $g$ matrix by introducing 
spin- and isospin-dependent multiplicative factors to simulate 
the 3NF effects. 
Results have shown that the 3NF effects are small for NA scattering, 
because the process is mainly governed by the interaction at low density 
regions. 
This reinforces the success of calculations using the Melbourne $g$ matrix 
without considering 3NFs. 
On the other hand, sizable effects are found for AA scattering at 
around 80 MeV/nucleon through the repulsive contribution in the real 
part and the enhanced absorptive potential. 
However, these are rather exploratory, relying on the 
Melbourne $g$ matrix. 
In this paper, we present full chiral $g$ matrices 
parameterized in a
3-range Gaussian form on the basis of nuclear matter $g$-matrix 
calculations with the 2NF and 3NF of Ch-EFT, and apply them to 
the SF and DF models.

The $g$-matrix DF model for AA scattering has a practical problem. 
In nuclear matter calculations, we have to consider two Fermi spheres and 
the $g$ matrix should be obtained by solving 
scattering between a nucleon in a Fermi sphere and a nucleon 
in another Fermi sphere~\cite{Izu80,Yahiro-Glauber}, 
but it is quite difficult in practice. 
In fact, the $g$ matrix is evaluated by solving nucleon scattering 
from a single Fermi sphere. 
For consistency with the nuclear-matter calculation, 
we assumed $\rho=\rho_{\rm T}(\vrr_{\rm m})$ in $g(\rho)$ 
and applied this framework to $^{3,4}$He scattering from heavier targets 
in a wide range of incident energies from 30~MeV/nucleon 
to 180~MeV/nucleon \cite{Egashira:2014zda,Toyokawa:2015fva}. 
The Melbourne $g$-matrix DF model with the target-density approximation (TDA) 
well reproduces the data with no adjustable parameter, 
particularly for total reaction cross sections 
$\sigma_{\rm R}$ and forward differential cross sections. 
The DF-TDA model does not include 
projectile-excitation effects, but it was confirmed by CDCC calculations 
that the effects are negligible for $^{3}$He scattering. Precisely, 
the effects are appreciable at incident energies lower than 40~MeV/nucleon, 
but they enhance $\sigma_{\rm R}$ only by a few percent. 
It is very likely that projectile-excitation effects 
are even smaller for $^{4}$He scattering, since $^{4}$He is less fragile  
than $^{3}$He. The practical problem is thus solved 
for $^{3,4}$He scattering. 
Therefore, this DF-TDA model is used in this paper.

For heavier projectiles than $^{4}$He, it is quite difficult to 
include all projectile-excitation effects explicitly. 
For such AA scattering, the frozen-density approximation (FDA), 
$\rho={\rho_{\rm P}(\vrr_{\rm m})}+{\rho_{\rm T}(\vrr_{\rm m})}$, 
is often taken as a value of $\rho$ in $g(\rho)$, although 
$g(\rho)$ is obtained by solving nucleon scattering 
on a single Fermi sphere. 
The DF-FDA model includes projectile-excitation effects approximately. 
The model based on the Melbourne $g$ matrix 
well reproduces measured $\sigma_{\rm R}$ 
for $^{12,14-16}$C \cite{Sasabe:2013dwa,Matsumoto:2014qda}
and Ne and Mg isotopes 
\cite{Sum12,Watanabe:2014zea}. 
As an important result, the microscopic analyses 
conclude that $^{31}$Ne and $^{37}$Mg are deformed halo nuclei. 
For $^{3,4}$He scattering, however, 
the DF-TDA model always yields 
better agreement with the experimental data than the DF-FDA model   
\cite{Egashira:2014zda,Toyokawa:2015fva}.

The $g$ matrix obtained is quite inconvenient in many applications, 
since it is nonlocal and numerical. 
The Melbourne group showed that elastic scattering are mainly determined by 
the on-shell part of $g(\rho)$ \cite{Amos}. 
Making a $\chi^2$ fitting to the on-shell and near-on-shell 
components of the $g$ matrix, 
the group provided $g(\rho)$ 
with a local (Yukawa) form 
in order to make the folding procedure feasible 
\cite{von-Geramb-1991,Amos-1994,Amos}. 
The Melbourne $g$ matrix thus obtained accounts for NN scattering 
in the limit of $\rho=0$, 
and the SF model based on the Melbourne $g$ matrix 
explains NA scattering systematically with no adjustable parameter, 
as mentioned above.

In this paper, we consider 
heavier targets such as $^{40}$Ca, $^{58}$Ni and $^{208}$Pb 
to make our discussion clear, since the $g$ matrix is evaluated 
in nuclear matter and the $g$-matrix folding model is 
considered to be more reliable for heavier targets.  
Taking the Melbourne-group procedure 
\cite{von-Geramb-1991,Amos-1994,Amos}, we provide 
the chiral $g$ matrix with a 3-range Gaussian form 
for each of the central, spin-orbit and tensor components, 
since the Gaussian form is much more convenient 
than the Yukawa form in many applications whereas 
the two forms yield the same results 
for NA and AA scattering. 
The ranges and the depths of individual components are determined 
for each energy and density 
so as to reproduce the on-shell and near-on-shell matrix elements 
of the original $g$ matrix. 
For the central part of $g$ matrix, 
the present ranges of 3-range Gaussian form
are $(0.4,0.9,2.5)$ fm 
and close to those of Ref.~\cite{CEG07}. 
We call the analytic form ``Gaussian chiral $g$ matrix" and 
the original numerical $g$ matrix ``original chiral $g$ matrix", 
when we need to identify the two. 
The folding model based on Gaussian chiral $g$ matrix 
reproduces the experimental data with no adjustable parameter 
for the present scattering. Therefore, 
we can investigate chiral-3NF effects on proton and $^{4}$He 
scattering clearly. 

In Sec. \ref{Theoretical framework}, we recapitulate 
the BHF method for the symmetric nuclear matter with 2NF+3NF
and the folding model for proton and $^{4}$He scattering.
In Sec. \ref{Results}, the results of 
the folding model with Gaussian chiral g matrix are shown 
for proton and $^{4}$He scattering.
Section \ref{Summary} is devoted to a summary.

\section{Theoretical framework}
\label{Theoretical framework}
\subsection{Nuclear-matter calculations for 3NF}
\label{Nuclear-matter calculations for 3NF}
We recapitulate the BHF method for the case of 2NF+3NF
\cite{Koh13}. 
The 3NF $V_{123}$ is hard to treat even in infinite matter. 
We then derive an effective 2NF $V_{12(3)}$ from $V_{123}$ 
by averaging it over the third nucleon in the Fermi sea. 
After this approximation, the potential energy is reduced to 
\begin{eqnarray}
 &&\frac{1}{2} \sum_{\vik_1 \vik_2} \langle \bk_1 \bk_2 |
  V_{12} |\bk_1 \bk_2\rangle_A 
  \nonumber\\
 &&~~~ +\frac{1}{3!}\sum_{\vik_1 \vik_2 \vik_3} \langle \bk_1 \bk_2 \bk_3| 
  V_{123} |\bk_1 \bk_2 \bk_3\rangle_A
  \nonumber \\
 &&~ =\frac{1}{2} \sum_{\vik_1 \vik_2} \langle \bk_1 \bk_2 | 
  V_{12}^{\rm eff}
|\bk_1 \bk_2\rangle_A  
\label{pot-energy}
\end{eqnarray}
with the effective 2NF
\bea
 V_{12}^{\rm eff}=V_{12}+ \frac{1}{3}V_{12(3)}, 
\label{eff-V12}
\eea
where the symbol $A$ means the antisymmetrization and 
$\bk_i$ denotes quantum numbers of the $i$-th nucleon; 
note the factor $1/3$ in front of $V_{12(3)}$ 
in Eq. \eqref{eff-V12}.  
The $g$ matrix $g_{12}$ is then obtained by solving 
the equation 
\begin{equation}
 g_{12}=V_{12}^{\rm eff}+V_{12}^{\rm eff} G_0 g_{12} 
\label{g-eq}
\end{equation}
for $g_{12}$ with the nucleon propagator $G_0$ including 
the Pauli exclusion operator. 
Here the single-particle energy $e_{\bk}$ 
for a nucleon with momentum ${\bk}$ in the denominator of $G_0$ 
is obtained by~\cite{Koh13}
\bea
e_{\vik}=\langle \bk |T|\bk\rangle + {\rm Re}[{\cal U}(\bk)]  
\eea
with the single-particle potential 
\bea
{\cal U}(\bk) = \sum_{\vik'}^{k_{\rm F}^{}} \langle \bk \bk' | g_{12} 
+\frac{1}{6} V_{12(3)}(1+G_0 g_{12}) 
|\bk \bk' \rangle_A ,
\label{single-particle-pot-0}
\eea
where $T$ is a kinetic-energy operator of nucleon with 
the mass $m$ and $\vk$ is related to 
the incident energy $E_{\rm in }$ as 
$E_{\rm in }=(\hbar \vk)^2/(2m) + {\rm Re}[{\cal U}]$. 
When $E_{\rm in }>0$, the single-particle potential 
is nothing but an optical potential 
of an extra nucleon in nuclear matter.
Similar calculations, but in the second-order perturbation, 
of the optical potential in the framework of Ch-EFT was
reported by Holt \textit{et al}. \cite{Hol13}.
The present formulation is consistent with theirs in virtue of 
the factor $1/6$ in Eq. \eqref{single-particle-pot-0} 
\cite{Koh13}.

In the present BHF calculation, the cutoff energy 
$\Lambda=550$ MeV is used with the form factor 
$\exp\{-(q'/\Lambda)^6-(q/\Lambda)^6\}$ both for 
$V_{12}$ and $V_{12(3)}$.
The low-energy constants of chiral forces are taken from Ref.~\cite{Epe05} as 
$(c_1,c_3,c_4)=(-0.81,-3.4,3.4)$ in units of GeV$^{-1}$, 
and the other constants $(c_D,c_E)=(-4.381,-1.126)$ 
are from Ref.~\cite{HEB11}. 
Other sets of low-energy constants present in literature
\cite{Entem-2003} are expected to give essentially same results.
Furthermore, the variation of $g$ matrices is much reduced 
in the effective 2NF level 
when 3NFs are incorporated consistently \cite{Koh13}. 
In addition, the net effect of $c_D$ and $c_E$ is small, 
when $c_D \simeq 4c_E$. 
This relation is well satisfied in various calculations 
for light nuclei in Ref.~\cite{Nogga-2006} and also for   
nuclear matter in Ref.~\cite{HEB11} and the present work. 
As for ${\cal U}$, our results are similar to 
those of second-order perturbation calculations \cite{Hol13} 
for the real part, but 
for the imaginary part the former is more absorptive 
than the latter. 
This may be originated in the full ladder-summation in $g$-matrix calculations.

\begin{figure}[tbp]
\begin{centering}
 \includegraphics[width=0.45\textwidth,clip]{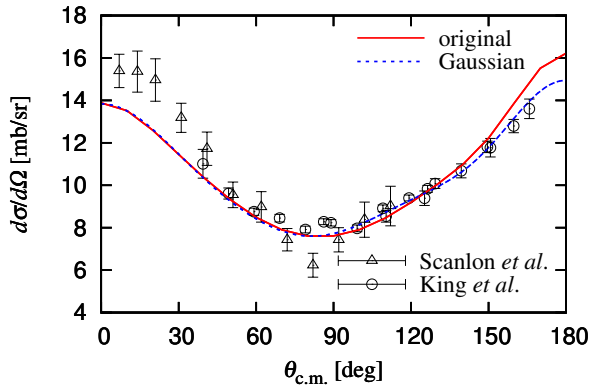}
 \caption{(Color online) 
Differential cross sections for neutron-proton scattering at 
$E_{\rm in } \simeq 65$ MeV in free space. 
The solid (dashed) curves represent the results of 
original (Gaussian) chiral $t$ matrix. 
Experimental data are taken from Refs.~\cite{np:Scanlon,np:King}.
}
 \label{fig:n+p-scattering}
\end{centering}
\end{figure}

Figure \ref{fig:n+p-scattering} shows differential cross sections 
for neutron-proton scattering at $E_{\rm in } \simeq 65$ MeV 
in free space 
(in the limit of $\rho=0$), 
where $E_{\rm in}$ stands for an incident energy 
in the laboratory system. 
The solid and dashed lines denote the results of 
original and Gaussian chiral $t$ matrices, respectively; 
note that the $g$ matrix is reduced to the $t$ matrix 
in the limit of $\rho=0$.  
Thus the Gaussian $t$ matrix well reproduces the result 
of original chiral $t$ matrix.

The $g$ matrix can be classified with $S, T, E_{\rm in }$ and 
$k_{\rm F}$ as $g^{ST}(k_{\rm F},E_{\rm in })$. 
Hence ${\cal U}$ can be decomposed into 
${\cal U}=\sum_{ST}(2S+1)(2T+1){\cal U}^{ST}$ with 
${\cal U}^{ST}$ defined by Eq. \eqref{single-particle-pot-0} 
in which $g$ and $V_{12(3)}$ are replaced by 
$g^{ST}$ and $V_{12(3)}^{ST}$, respectively. 
Thus, ${\cal U}^{ST}$ means the 
single-particle potential 
in each spin-isospin channel.

\begin{figure}[tbp]
\begin{centering}
 \includegraphics[width=0.48\textwidth]{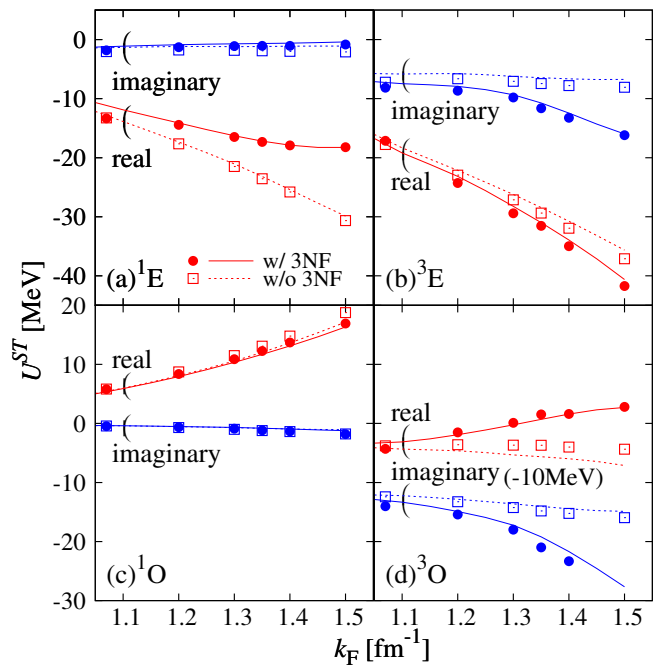}
\caption{(Color online) 
$k_{\rm F}$ dependence of ${\cal U}^{ST}$ 
at $E_{\rm in}=65$ MeV  
for (a)$^{1}$E, (b) $^{3}$E, (c) $^{1}$O, and (d) $^{3}$O. 
Squares (circles) mean the results of original chiral $g$ matrix with 
(without) 3NF. 
The solid (dashed) lines represent the results of 
Gaussian chiral 
$g$ matrix with (without) 3NF. 
For $^{3}$O, the imaginary part is shifted down by 10 MeV. 
}
\label{fig-SPP-ST}
\end{centering}
\end{figure}

Figure \ref{fig-SPP-ST} shows  $k_{\rm F}$ dependence of ${\cal U}^{ST}$. 
The squares and circles stand for the results of 
original chiral $g$ matrix 
with and without 
chiral 3NF, respectively. The difference between the two results 
mainly stems from the 2$\pi$-exchange diagram in chiral 3NF. 
Particularly for nucleon and $^{4}$He scattering, 
the region 
$k_{\rm F} \la 1.35$ fm$^{-1}$ ($\rho \la \rho_0$) is important. 
For the spin-triplet channels ($^{3}$E and $^{3}$O), 
the 2$\pi$-exchange 3NF enhances tensor correlations and makes 
transitions between different states stronger, and eventually it 
makes the imaginary part of ${\cal U}^{ST}$ more absorptive. 
For $^{1}$E, chiral-3NF effects are large and 
repulsive, which corresponds to the suppression of $\Delta$ isobar 
excitations in nuclear medium in a conventional picture. 
The solid and dashed lines correspond to 
the results of Gaussian chiral $g$ matrix with and without 
chiral 3NF, respectively. 
The Gaussian chiral $g$ matrix well 
reproduces the results of the original chiral $g$ matrix. 
The potentials in the parity-odd channels, $^{1}$O and $^{3}$O, 
are small.

\subsection{Folding model} 
\label{Folding model} 
The formulation of SF and DF models is summarized 
in Ref.~\cite{Toyokawa:2015fva}, 
together with the relation between the two models. 
The folding potential is composed 
of the direct and knock-on exchange components. 
The latter component makes the potential nonlocal, 
but it can be localized with high accuracy by 
the Brieva-Rook (local momentum) approximation \cite{Brieva-Rook}. 
The reliability of this approximation is shown in 
Refs.~\cite{Hag06,Minomo:2009ds}. 
The resultant folding potential 
$U(\vR)$ is a function of the relative coordinate $\vR$ of 
projectile and target. 
The odd ($^{3}$O and $^{1}$O) channels of $g^{ST}$ 
are almost canceled between the direct and knock-on exchange components, 
and hence $U(\vR)$ is determined mainly by 
the even ($^{3}$E and $^{1}$E) $g$ matrices; e.g., 
see Refs.~\cite{Minomo:2014eqa,Toyokawa:2014yma}. 
The $S$ matrices for 
NA and $^{4}$He elastic scattering are obtained by solving 
a one-body Schr\"odinger equation with $U(\vR)$.

For $^{40}$Ca, $^{58}$Ni and $^{208}$Pb targets, 
the matter densities are calculated with 
the spherical Hartree-Fock (HF) method using  
the Gogny-D1S interaction~\cite{GognyD1S}.
The spurious center-of-mass (c.m.) motions are 
removed in a standard prescription~\cite{Sum12}. 
For $^{4}$He,  we take 
phenomenological density determined 
from electron scattering~\cite{phen-density} in which 
the finite-size effect of proton charge 
is unfolded 
by using a standard procedure~\cite{Singhal}.

\section{Results}
\label{Results}

\begin{figure}[tbp]
 \begin{centering}
  \includegraphics[width=0.45\textwidth,clip]{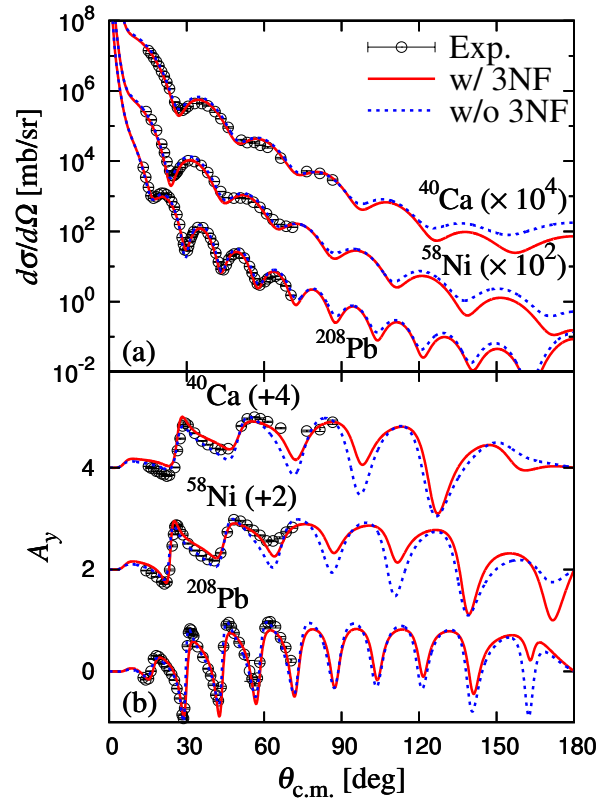}  
 \caption{(Color online) 
Angular distribution of (a) differential cross sections and 
(b) vector analyzing powers for 
proton elastic scattering at 65 MeV. 
The solid (dashed) lines denote the results of 
chiral $g$ matrix with (without) 3NF effects. 
Each cross section is multiplied by the factor shown in the figure, 
while each vector analyzing power is shifted up by the number shown 
in the figure. 
Experimental data are taken from Ref.~\cite{pCa40:Ni58:Pb208:Sakaguchi}.
}
 \label{fig:p+A-scattering}
 \end{centering}
\end{figure}

First, we consider proton elastic scattering at $E_{\rm in}=65$ MeV 
from $^{40}$Ca, $^{58}$Ni and $^{208}$Pb targets.
In Fig.~\ref{fig:p+A-scattering}, 
differential cross sections $d \sigma/d \Omega$ and 
vector analyzing powers $A_y$ are plotted as 
a function of scattering angle $\theta_{\rm c.m.}$ in the c.m. system. 
The solid and dashed lines stand for the results of 
chiral $g$ matrix with and without 3NF effects, respectively. 
Chiral-3NF effects are small 
at forward and middle angles 
where the experimental data 
\cite{pCa40:Ni58:Pb208:Sakaguchi} are available,  
since the scattering is governed by the potential 
in the surface region where 3NF effects are small because of low density. 
Only an exception is $A_y$ at 
$\theta_{\rm c.m.} \simeq 60^\circ$ 
for $^{40}$Ca and $^{58}$Ni targets. 
Chiral-3NF effects enhance the spin-orbit part 
of $U(\vR)$ by a factor of about 30~\%, 
which may be the reason for this improvement. 
We confirmed that chiral-3NF effects are small also 
for $\sigma_{\rm R}$.

\begin{figure}[tbp]
\begin{centering}
 \includegraphics[width=0.45\textwidth,clip]{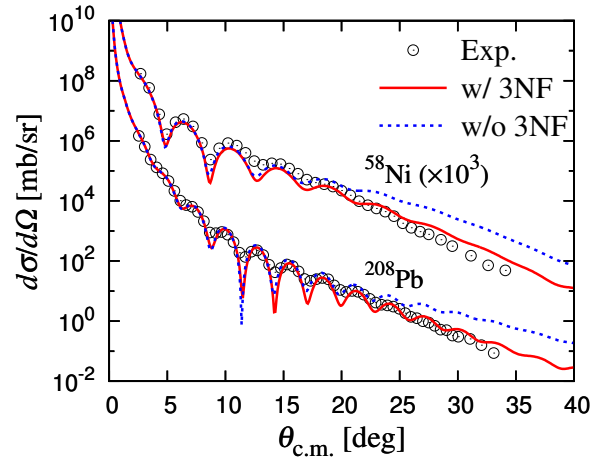}
 \caption{(Color online) 
Angular distribution of differential cross sections for 
$^{4}$He scattering at 72MeV/nucleon 
from $^{58}$Ni and $^{208}$Pb targets. 
The solid (dashed) lines denote the results of 
chiral $g$ matrix with (without) 3NF effects.
Each cross section is multiplied by the factor shown in the figure.  
Experimental data are taken from Ref.~\cite{He4Ni58:Pb208:Bonin}. 
}
 \label{fig:4He+A-scattering}
\end{centering}
\end{figure}

Next, we show the angular distribution of $d \sigma/d \Omega$
for $^{4}$He scattering at 72MeV/nucleon 
from $^{58}$Ni and $^{208}$Pb targets in Fig. \ref{fig:4He+A-scattering}. 
The solid and dashed lines denote the results of 
chiral $g$ matrix with and without 3NF effects, respectively. 
For both targets, chiral-3NF effects are sizable at middle angles 
$\theta_{\rm c.m.} \ga 20^\circ$ where the experimental data 
\cite{He4Ni58:Pb208:Bonin}
are available.

The scattering amplitude can be decomposed into the near- 
and far-side components \cite{Fuller:1975}. 
When a detector is set on the right-hand side of the target, 
the outgoing wave going through the right-hand (left-hand) side 
of the target is called the near-side (far-side) scattering. 
The near-side (far-side) component is mainly induced by 
repulsive Coulomb (attractive nuclear) force, 
and in general the near-side (far-side) component dominates 
forward-angle (middle-angle) scattering. 
For both $^{58}$Ni and $^{208}$Pb targets, 
large oscillations seen in the range 
$\theta_{\rm c.m.}= 5$ -- $20^\circ$ are 
a consequence of the interference between 
the near- and far-side components. 
When the scattering is dominated by the far-side component, 
$d \sigma/d \Omega$ has no oscillation and is sensitive 
to the change of nuclear force. 
The far-side dominance appears at $\theta_{\rm c.m.} > 20^\circ$. 
Chiral-3NF effects thus appear in the far-side dominant 
angles sensitive to the change of nuclear force.

\begin{figure}[htbp]
\begin{centering}
 \includegraphics[width=0.38\textwidth,clip]{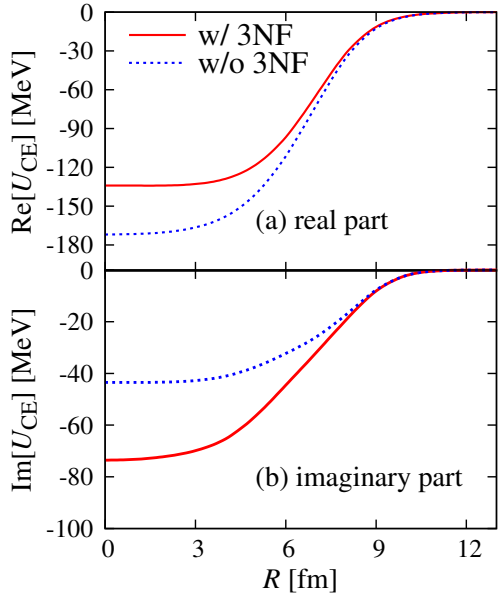}
 \caption{(Color online) 
$R$ dependence of the central part of the folding potential 
for $^{4}$He+$^{208}$Pb elastic scattering at $E=72$~MeV/nucleon. 
Panels (a) and (b) correspond to the real and 
imaginary parts of $U_{\rm CE}(R)$, respectively. 
The solid (dashed) lines represent the results of 
chiral $g$ matrix with (without) chiral 3NF. 
}
\label{Fig:pot;4He+Ni;E=72}
\end{centering}
\end{figure}

Figure \ref{Fig:pot;4He+Ni;E=72} shows the central part 
$U_{\rm CE}(R)$  of $U$ 
for $^{4}$He + $^{208}$Pb scattering at 72MeV/nucleon. 
The solid and dashed lines correspond to the results of 
chiral $g$ matrix with and without 3NF effects, respectively. 
Chiral 3NF, mainly in its the 2$\pi$-exchange diagram, 
makes $U_{\rm CE}(R)$ less attractive and more absorptive. 
This repulsive effect of chiral 3NF in 
$U_{\rm CE}(R)$ comes from the repulsion 
in the $^{1}$E channel of $g^{ST}$. 
The repulsive nature suppresses $d \sigma/d \Omega$ 
at $\theta_{\rm c.m.} > 20^\circ$ for $^{4}$He scattering, 
whereas stronger absorption due to chiral 3NF better separates 
the far-side amplitude from the near-side one.

\section{Summary} 
\label{Summary} 
We investigated the effects of chiral NNLO 3NF 
on proton scattering at 65 MeV and $^{4}$He scattering 
at 72 MeV/nucleon from heavier targets, 
using the standard 
BHF method and the $g$-matrix folding model. 
We evaluated the $g$ matrix from N$^{3}$LO 2NF plus NNLO 3NF for 
positive energy 
in nuclear matter.
The same calculations for negative energies account well for the empirical 
saturation properties of symmetric nuclear matter. 
Chiral-3NF effects are mainly originated in 
the 2$\pi$-exchange diagram. 
The 3NF contribution in the $^{3}$E channel enhances tensor correlations
to make the optical potential more absorptive. 
In the $^{1}$E channel, the 3NF effect yields a repulsion that may
correspond to the Pauli suppression of isobar $\Delta$ excitation 
in the nuclear-matter medium in the conventional picture.

Following the Melbourne-group procedure 
\cite{von-Geramb-1991,Amos-1994,Amos}, we provided the chiral $g$ matrix 
with a 3-range Gaussian form by making a $\chi^2$ fitting to 
the on-shell and near-on-shell parts of the original numerical 
$g$ matrix. This Gaussian form makes the folding procedure much easier. 
The $g$-matrix folding model with chiral 3NF reproduces 
the experimental data with no adjustable parameter for proton and 
$^{4}$He scattering.  
We found that chiral-3NF effects are small 
for proton scattering but sizable for $^{4}$He scattering at 
middle angles $\theta_{\rm c.m.} \ga 20^\circ$ 
where the experimental data are available. 
Chiral 3NF
yields repulsive and absorptive corrections to $U_{\rm CE}(R)$ 
for both proton and $^{4}$He scattering. 
$^{4}$He scattering is dominated by the far-side 
scattering amplitude at middle angles $\theta_{\rm c.m.} \ga 20^\circ$. 
The repulsive nature of chiral 3NF 
suppresses the far-side scattering amplitude, 
whereas the absorptive nature of chiral 3NF better 
separates the far-side scattering from the near-side one. 
Chiral 3NF thus becomes sizable in the far-side dominant angle range. 
Note that the repulsive nature comes from 
the $^{1}$E channel, whereas the absorptive nature 
from the $^{3}$E channel.
Phenomenological 3NFs also make repulsive corrections 
to $U_{\rm CE}(R)$ ~\cite{CEG07,Raf13,MP}. 
However, the origin of the repulsion is different. 
It is interesting if the mechanism of producing the repulsive
contributions is clarified by analyzing scattering data.

\section*{Acknowledgements} 
This work is supported in part by
by Grant-in-Aid for Scientific Research
(Nos. 25400266, and 26400278)
from Japan Society for the Promotion of Science (JSPS).


\end{document}